\documentclass[seceq,supplement]{ptptex}
\title{
Time correlation function of the shear stress 
in sheared particle systems%
}

\author{
  Michio \textsc{Otsuki} and Hisao \textsc{Hayakawa}%
}

\inst{
Yukawa Institute for Theoretical Physics, Kyoto University,  Kitashirakawaoiwake-cho, Sakyo,
Kyoto 606-8502, JAPAN
}
\usepackage{amsmath}
\usepackage{dcolumn}
\usepackage[xdvi]{graphicx}%
\makeatletter

\def\btt#1{\texttt{\@backslashchar#1}}%
\DeclareRobustCommand\bblash{\btt{\@backslashchar}}%

\newcommand{\bv}[1]{{\boldsymbol #1}}

\abst{
The long-time behaviors of the time correlation function $C_\eta(t)$ 
of the shear stress
for three-dimensional sheared fluids
are investigated theoretically. 
It is found that there are the cross-overs in
$C_\eta(t)$ from $t^{-3/2}$ to $t^{-2}$  
for sheared fluids of elastic particles without any thermostat,  and
from $t^{-3/2}$ to $t^{-5/2}$ 
for isothermal sheared fluids including granular fluids.
}

\begin{document}

\maketitle



\section{Introduction}

The long time behaviors of current autocorrelation functions are
important to understand the macroscopic properties of fluids 
\cite{alder,ernst71,Pomeau,Dorfman}.
In is known that the existence of the long-time tail in the correlation
functions leads to the anomalous behaviors of the transport coefficients 
\cite{Narayan,Murakami}. 

In equilibrium systems, the existence of the long-time tail 
$t^{-d/2}$ with the time $t$ and the spatial dimension $d$ is well recognized.
However, the long time behaviors
of the correlation functions under a steady shear have different
feature from those at equilibrium \cite{Kumaran, Otsuki07}.
In our previous paper, we find that the velocity autocorrelation 
function $C(t)$ has the cross-over from $t^{-d/2}$ to $t^{-d}$ in sheared
elastic particles without thermostat, and 
$C(t)$ obeys $t^{-(d+2)/2}$ after the known tail $t^{-d/2}$ in sheared
isothermal fluids \cite{Otsuki07}.
However, we did not discuss the other correlation functions
such as the correlation of the shear stress and the related transport
coefficients.

In this paper, thus,  we theoretically calculate 
the correlation function of the shear stress. 
Our theoretical method is based on the classical one developed by Ernst 
\textit{et al}. \cite{ernst71,Hayakawa}.
In the next section, we will introduce the model we use.  In section 3, we will present the details of the analysis.
In section 4, we will discuss and conclude our results.

\section{Model}

We consider
a system consists of $N$ identical smooth and hard spherical particles with the mass
$m$ and 
the diameter $\sigma$ in the volume $V$. 
The position and the velocity of the $i$-th particle at time $t$
are $\bv{r}_i(t)$ and $\bv{v}_i(t)$, respectively.  
The particles collide instantaneously 
with the restitution constant $e$ which is equal to unity for elastic particles
or is less than unity for granular particles.
When the particle $i$ with the velocity $\bv{v}_i$ collides with the particle  $j$ with $\bv{v}_j$,
the post-collisional velocities $\bv{v}'_i$ and $\bv{v}'_j$ are respectively given by
$\bv{v}'_i  =  \bv{v}_i  - \frac{1}{2} (1+e) (\bv{\epsilon} \cdot
\bv{v}_{ij}) \bv{\epsilon} $ and
$\bv{v}'_j  =  \bv{v}_j  + \frac{1}{2} (1+e) (\bv{\epsilon} \cdot
\bv{v}_{ij})\bv{\epsilon} $,
where $\bv{\epsilon}$ is the unit vector parallel to the relative position of the
two colliding particles at contact, and $\bv{v}_{ij}=\bv{v}_{i}-\bv{v}_{j}$. 

Let us assume that the uniform shear flow is stable and
its velocity profile is given by
$c_{\alpha}(\bv{r}) = \dot{\gamma} y \delta_{\alpha,x}$,
where the Greek suffix $\alpha$ denotes the Cartesian component.
In this paper, we discuss the following situations: (a) A sheared system 
of elastic particles
without any thermostat, (b)  a sheared system of elastic particles with 
the velocity rescaling thermostat, and (c) 
a sheared granular system with the restitution constant $e<1$.
We abbreviate them to the sheared heating (SH),  the sheared thermostat (ST),  
and the sheared granular (SG) systems for later discussion.

We are interested in the correlation function of the shear stress,
\begin{eqnarray}
C_\eta(t) &=& \lim_{V\to\infty}\frac{1}{V}\langle J_{\eta}(0)
J_{\eta}(t) \rangle, \label{eq4}
\end{eqnarray}
where $J_\eta(t)$ is  the shear stress
at time $t$. Here,  $t=0$ is the time when we start the measurement. 
In general,  the current $J_{\eta}(t)$ consist of the kinetic part and the potential part.
In this paper, we only consider the contribution from the kinetic part of the current. This treatment is correct
for dilute gases. For higher density cases, we need more sophisticated method to include the contribution from
the potential part, but the corrections only appear in the prefactor of coefficients at least for elastic gases in the equilibrium state 
\cite{dorfman75,ernst76}.
Thus, the currents in Eq. (\ref{eq4}) is approximated  
by the kinetic part $J_{\eta}^K(t)$ as
\begin{eqnarray}
J_{\eta}(t) & \simeq & J_{\eta}^K(t)\equiv \sum_i m v'_{ix}(t)v'_{iy}(t),
 \label{eq7}
\end{eqnarray}
where $\bv{v}'_i(t)$ is the peculiar velocity defined as
 $\bv{v}'_i(t)= \bv{v}_i(t) - \bv{c}(\bv{r}_i(t))$.

\section{Theoretical analysis}

\subsection{Hydrodynamic equations}

Following Ernst {\it et al.} \cite{ernst71}, 
we approximate the correlation function  (\ref{eq4}) by
\begin{eqnarray}
C_\eta(t) &\simeq&  m^2 n_H \int d\bv{v}'_0
  v'_{0x}v'_{0y}
 f_0(v'_0) \int d \bv{r}   u_x(\bv{r},t)  (u_y(\bv{r},t)-\dot{\gamma} y) ,
 \label{C2:eq}
\end{eqnarray}
where $f_0(v'_0)$ and $\bv{u}(\bv{r},t)$ are the distribution function 
of the peculiar velocity at the initial time $t=0$, 
the velocity field at the position 
$\bv{r}$ and the time $t$, respectively.

The velocity field $\bv{u}(\bv{r},t)$ obeys the hydrodynamic equations as \cite{Brey}
\begin{eqnarray}
\partial_t n + \bv{\nabla} \cdot (n\bv{u}) & = & 0, \nonumber \\
\partial_t \bv{u} + \bv{u} \cdot \bv{\nabla} \bv{u} 
+ (nm)^{-1} \bv{\nabla} \cdot \bv{\Pi}& = & 0,  \nonumber \\
\partial_t T + \bv{u} \cdot \bv{\nabla} T 
+ 2 (dn)^{-1} (\bv{\Pi}:\bv{\nabla}\bv{u} 
  -\kappa \nabla^2  T - \mu \nabla^2 n)
+T \zeta & = & 0, \label{Heq}
\end{eqnarray}
where $n(\bv{r},t)$ and $T(\bv{r},t)$ are the density and the temperature
field, respectively. 
The pressure tensor $\bv{\Pi}$ is given by
\begin{eqnarray}
\Pi_{ij} & = & nT \delta_{ij} - \eta \left (\nabla_i u_j + \nabla_j u_i - 
\frac{2}{d} \delta_{ij} \nabla_k u_k \right ). \label{Pi}
\end{eqnarray}
Note that the bulk viscosity is zero for fluids of dilute hard 
spheres.
$\zeta$, $\lambda$, $\mu$, and $\eta$
are the cooling rate,
the heat conductivity, and the transport coefficient associated with the
density gradient, and the viscosity, respectively.
Here, $\mu$ has a finite value when $e$ is less than unity
but becomes zero in the elastic case.

Here, the viscosity $\eta$, the heat conductivity $\lambda$, and the transport coefficient associated with the density gradient $\mu$
can be non-dimensionalized as
\begin{eqnarray}
\eta  =  \eta_0 \eta^*, \qquad
\kappa  =  \kappa_0 \kappa^*, \qquad
\mu  =  \frac{T\kappa_0}{n} \mu^*,
\end{eqnarray}
where
\begin{eqnarray}
\eta_0  =  a \sqrt{T},  \qquad
\kappa_0  =  \frac{d(d+2)}{2m(d-1)} a \sqrt{T}
\end{eqnarray}
are the viscosity and the heat conductivity in the dilute gas \cite{Brey}. 
Here, $\eta^*$, $\kappa^*$, and $\mu^*$ are the constants 
depend only on $e$ in dilute cases. 
Here, the explicit form of $a$ is given by
$a = \frac{2+d}{8} \Gamma(d/2) \pi ^{-\frac{d-1}{2}} \sqrt{m}
\sigma^{-(d-1)}$ with the gamma function $\Gamma(x)$.

It is obvious that there is the relation 
$\zeta  =  0$ for SH. 
The cooling rate $\zeta$ becomes 
\begin{eqnarray}
\zeta & = & \frac{2 \eta(T) \dot{\gamma}^2}{d n_H T},
\label{zeta:thermo}
\end{eqnarray}
for ST.
On the other hand, the cooling rate is represented by
\begin{eqnarray}
\zeta & = & \frac{nT}{\eta_0} \zeta^*
\end{eqnarray}
for SG,
where $\zeta^*$ is the constants
proportional to $1-e^2$.

\subsection{fluctuation of the hydrodynamic fields}

The hydrodynamic equations (\ref{Heq}) have the set of homogeneous solutions as
\begin{eqnarray}
n(\bv{r}, t)  =  n_H, \qquad
u_{\alpha}(\bv{r}, t)  =  \dot\gamma y\delta_{\alpha,x}, \qquad
T(\bv{r}, t)  =  T_{H}(t), \label{Tdev}
\end{eqnarray}
where $n_H =N/V$ is the homogeneous density. The homogeneous
temperature $T_H(t)$ is 
\begin{eqnarray}
T_{H}(t) =  T_0
\left( 
1 + b  t
\right )^2, 
\label{THh}
\end{eqnarray}
where we introduce the initial temperature $T_0$,
the characteristic frequency $\nu_{H} = n_H T_H/\eta_0(T_H)$,
$\nu_{H0} = \nu_H(T_0)$ and $b = \eta^* \dot{\gamma}^2 / (3\nu_{H0})$
for SH. On the other hand, $T_H$ keeps a constant as
\begin{eqnarray}
T_{H}(t) =  T_0, \label{THt}
\end{eqnarray}
for ST and SG.
We can take any value as $T_0$ for ST, although 
$T_0$ depends on $\dot{\gamma}$ as
\begin{eqnarray}
T_{0} =  \frac{2a^2 \eta^*}{dn_H \zeta^*} \dot{\gamma}^2 \label{THd}
\end{eqnarray}
for SG.

Let us introduce the Fourier transform of any hydrodynamic field $f(\bv{r},t)$ is defined as
\begin{eqnarray}
\tilde{f}(\bv{k},T) = \int d \bv{\xi} \exp(-\bv{k} \cdot \bv{\xi}) 
f(\bv{r},T),
\label{Fourier}
\end{eqnarray}
where $\bv{\xi}$ is the non-dimensional position defined as
$\bv{\xi}  =  l_H^{-1} \bv{r}$,
with $l_H =2u_H/ \nu_{H0}$.

Let us write the Fourier transport $\tilde{\bv{u}}(\bv{k},t)$ of 
the peculiar velocity $\bv{u}(\bv{r},t) - \bv{c}(\bv{r})$ as
\begin{eqnarray}
\tilde{\bv{u}}(\bv{k},t) = \tilde{u}^{(1)}(\bv{k},t) \bv{e}^{(1)}(\bv{k}) +  
\tilde{u}^{(2)}(\bv{k},t) \bv{e}^{(2)}(\bv{k}) +  \tilde{u}^{(3)}(\bv{k},t) \bv{e}^{(3)}(\bv{k}),
\label{u3d:def}
\end{eqnarray}
where the unit vectors $\bv{e}^{(i)}(\bv{k})$
depending on $\bv{k}$ are defined as
\begin{eqnarray}
 \bv{e}^{(1)}(\bv{k}) & = & \hat{\bv{k}}  = (\hat{k}_x,\hat{k}_y,\hat{k}_z) \nonumber \\
 \bv{e}^{(2)}(\bv{k}) & = & 
 \frac{\bv{e}^y - ( \bv{e}^{(1)} \cdot \bv{e}^{(y)}) \bv{e}^{(1)} }
 {|\bv{e}^y - ( \bv{e}^{(1)} \cdot \bv{e}^{(y)}) \bv{e}^{(1)} |}   \nonumber \\
\bv{e}^{(3)}(\bv{k}) & = &  \bv{e}^{(1)}  \times \bv{e}^{(2)}.
\label{unit}
\end{eqnarray}
The $i$-th component of the velocity filed in Eq. (\ref{u3d:def}) is
expressed as $\tilde{u}^{(i)}(\bv{k},t) = \tilde{\bv{u}}(\bv{k},t) \cdot \bv{e}^{(i)}(\bv{k})$.

From the method used in the previous work \cite{Lutsko},
we can rewrite the velocity field as
\begin{eqnarray}
\tilde{u}^{(1)}(\bv{k},t)  & = &\frac{1}{4}u_{H0} B(t)
\left \{ \frac{\tilde{n}(\bv{k}(t),0)}{n_H} 
+ \frac{\tilde{T}(\bv{k}(t),0))}{T_0} \right \}
\left \{ E^{(1)}(\bv{k}(t),t) - E^{(2)}(\bv{k}(t),t) \right \}  \nonumber \\
& & + \frac{1}{2} B(t)\tilde{u}^{(1)}(\bv{k}(t),0) 
\left \{ E^{(1)}(\bv{k}(t),t) + E^{(2)}(\bv{k}(t),t)\right \}, \nonumber
\\
\tilde{u}^{(2)}(\bv{k},t)  & = &  B(t)\tilde{u}^{(2)}(\bv{k}(t),0) 
E^{(3)}(\bv{k}(t),t), \nonumber  \\
\tilde{u}^{(3)}(\bv{k},t)  & = &  B(t) \tilde{u}^{(3)}(\bv{k}(t),0) E^{(4)}(\bv{k}(t),t) +  B(t) \tilde{u}^{(2)}(\bv{k}(t),0) F(\bv{k}(t),t),
   \label{u:sol}
\end{eqnarray}
where we introduce
\begin{eqnarray}
\bv{k}(t) = (k_x, k_y-\dot{\gamma}t k_x,k_z),
\end{eqnarray}
\begin{eqnarray}
 B(t) = 
\left \{
\begin{array}{cc}
 \left( 
1 + b t
\right)
 & \mbox{for SH}, \\
 1
 & \mbox{for ST and SG}.\\
\end{array} 
\right. 
 \label{BSH}
\end{eqnarray}
In addition, we also introduce
\begin{eqnarray}
E^{(1)}(\bv{k},t) & = &\sqrt{\hat{k}(t)} e^{i \sqrt{2} k \alpha(t) 
- \frac{\Gamma_0}{4}\beta(t)k^2}, \nonumber \\
E^{(2)}(\bv{k},t) & = & \sqrt{\hat{k}(t)} e^{-i \sqrt{2} k \alpha(t) 
- \frac{\Gamma_0}{4}\beta(t)k^2}, \nonumber \\
E^{(3)}(\bv{k},t) & = & \frac{e^{ - \frac{\eta^*}{4} \beta(t)k^2}}{\hat{k}(-t)},
\nonumber \\
E^{(4)}(\bv{k},t) & = & e^{ - \frac{\eta^*}{4} \beta(t)k^2},
\label{E:def}
\end{eqnarray}
and
\begin{eqnarray}
F(\bv{k},t) = 
   M(\bv{k(-t)
   }) E^{(3)}(\bv{k},t) -   M(\bv{k}) E^{(4)}(\bv{k},t)
\end{eqnarray}
with
$\Gamma_0  =  \eta^*/2 + \mu^* + \kappa^*$, 
and $M(\bv{k}) = - k k_z / (k_x k_\perp) \tan^{-1}(k_y / k_\perp)$.

Here, we use 
\begin{eqnarray}
\alpha(t) & = &
\left \{
\begin{array}{cc}
\alpha_1(t) + \frac{\eta^* }{3} \left ( \frac{\dot{\gamma}}{\nu_{H0}} \right )^2 \alpha_2(t)
 & \mbox{for SH}\\
 \alpha_1(t) 
 & \mbox{for ST and SG},
\end{array} 
\right. 
\end{eqnarray}
and
\begin{eqnarray}
\beta(t) =  A_1(t) - A_2(t) \hat{k}_x \hat{k}_y
 + A_3(t) \hat{k}_x^2
\end{eqnarray}
with
\begin{eqnarray}
\alpha_1(t) & = & \frac{\nu_{H0}}{2\dot{\gamma} \hat{k}_x}
  (\hat{k}_y -\hat{k}_y(-t)\hat{k}(-t)) \nonumber \\
  & & - \frac{\nu_{H0}}{2\dot{\gamma} \hat{k}_x}(\hat{k}_y^2 -1) {\mathrm sgn}(\dot{\gamma}  \hat{k}_x)
  \log \left [ 
  \frac{|\hat{k}_y(-t) - {\mathrm sgn}(\dot{\gamma}  \hat{k}_x)\hat{k}(-t)|}
  {|\hat{k}_y - {\mathrm sgn}(\dot{\gamma}  \hat{k}_x)|}
  \right ], \\
  \alpha_2(t) & = & \frac{\nu_{H0}^2(\hat{k}(-t)^3 -1)}
  {\dot{\gamma}^2 \hat{k}_x^2}
  + \frac{\nu_{H0}\hat{k}_y}{\dot{\gamma} \hat{k}_x} \alpha_1(t),
\end{eqnarray}
and 
\begin{eqnarray}
A_1(t) & = & 
\left \{
\begin{array}{cc}
\nu_{H0} t \left ( 1 + \frac{\eta^*}{6} 
\left ( \frac{\dot{\gamma}}{\nu_{H0}} \right )^2 t \right ) & \mbox{for SH}\\
\nu_{H0} t  & \mbox{for ST and SG}, \\
\end{array} 
\right. 
\end{eqnarray}
\begin{eqnarray}
A_2(t) & = & 
\left \{
\begin{array}{cc}
\left ( \frac{\dot{\gamma}}{\nu_{H0}} \right ) 
(\nu_{H0} t)^2 
\left ( 1 + \frac{2\eta^*}{9} 
\left ( \frac{\dot{\gamma}}{\nu_{H0}} \right )^2 \nu_{H0} t \right ) &
\mbox{for SH} \\
\left ( \frac{\dot{\gamma}}{\nu_{H0}} \right ) 
(\nu_{H0} t)^2 & \mbox{for ST and SG}, \\
\end{array} 
\right. 
\end{eqnarray}
\begin{eqnarray}
A_3(t) & = & 
\left \{
\begin{array}{cc}
\frac{1}{3} 
\left ( \frac{\dot{\gamma}}{\nu_{H0}} \right )^2 (\nu_{H0} t)^3 
\left ( 1 + \frac{\eta^*}{4} 
\left ( \frac{\dot{\gamma}}{\nu_{H0}} \right )^2 \nu_{H0} t \right ) &
\mbox{for SH}, \\
\frac{1}{3} 
\left ( \frac{\dot{\gamma}}{\nu_{H0}} \right )^2 (\nu_{H0} t)^3 &
\mbox{for ST and SG}, \\
\end{array} 
\right. 
\end{eqnarray}
where $k = \sqrt{k_x^2 + k_y^2  + k_z^2}$, 
$k(t) = \sqrt{k_x^2 + (k_y - \dot{\gamma} t k_x)^2  + k_z^2}$,
$\hat{k}_\alpha = k_\alpha/k$, and $\hat{k}(t) = k(t)/k$ in Eq. (\ref{E:def})

\subsection{Result}

Substituting Eq. (\ref{u3d:def}) into (\ref{C2:eq})
with the definition of the Fourier transform (\ref{Fourier}),
the correlation function can be written as
\begin{eqnarray}
C_\eta(t) &=&  \sum_{i,j=1,2,3} C^{(ij)}_\eta(t),
\end{eqnarray}
where we introduce
\begin{eqnarray}
C^{(ij)}_\eta(t) & \equiv &  m^2 n_H l_H^d \int d\bv{v}'_0
  v'_{0x}v'_{0y} f_0(v'_0) \nonumber \\
  & & \times 
  \int \frac{d \bv{k}}{(2\pi)^d}   u^{(i)}(\bv{k},t)  u^{(j)}(-\bv{k},t)
 e^{(i)}_x(\bv{k}) e^{(j)}_y(\bv{k}).
 \label{Cij:def}
\end{eqnarray}
Then, the correlation function can be approximated as
\begin{eqnarray}
C_\eta(t) & \simeq &  C^{(11)}_\eta(t) +  C^{(22)}_\eta(t)  +  C^{(32)}_\eta(t) . \label{Ceq:exp}
\end{eqnarray}
The reason to neglect $C^{(12)}_\eta(t)$,  $C^{(21)}_\eta(t)$, 
and  $C^{(31)}_\eta(t)$ is that these functions decay faster than
 $C^{(22)}_\eta(t)$ because of the existence of
the sound mode \cite{ernst71}.
The reason to neglect $C^{(13)}_\eta(t)$ and $C^{(23)}_\eta(t)$
is that $e^{(3)}_y(\bv{k})$ is zero.

In order to obtain $C_\eta^{(11)}(t)$, $C_\eta^{(22)}(t)$, 
and $C_\eta^{(32)}(t)$,
we use 
$\tilde{\bv{u}}_{\bv{k}}(0) \simeq \bv{v}_0/(n_0 l_H^d)$, 
and
$f_0(\bv{v}'_0) \simeq n (m/2\pi T_0)^{3/2} \exp(-c^2)\{1 + a_2 S_2(c^2)\}$,
where $c=v'\sqrt{ m/2T_0}$, $S_2(x) = x^2/2 - 5 x/2 + 15/8$, and
$a_2$ is the constant depending on the restitution coefficient $e$
\cite{ernst71,Hayakawa}. 
Here, we note that $a_2$ becomes zero when $e=1$.
Then , $C_\eta^{(11)}(t)$, $C_\eta^{(22)}(t)$, and $C_\eta^{(32)}(t)$,
are respectively calculated as
\begin{eqnarray}
C_\eta^{(11)}(t) &=&  \frac{T_0^2(1+a_2)}{8\pi^3 l_H^d}
\left (  
\frac{2}{\Gamma_0}
\right )^{3/2}
\psi^{(11)}(t), \nonumber \\
C_\eta^{(22)}(t) &=&  \frac{T_0^2(1+a_2)}{4 \pi^3 l_H^d} 
\left (  
\frac{2}{\eta^*}
\right )^{3/2}
\psi^{(22)}(t), \nonumber \\
C_\eta^{(32)}(t) &=&  \frac{T_0^2(1+a_2)}{4 \pi^3 l_H^d} 
\left (  
\frac{2}{\eta^*}
\right )^{3/2}
\psi^{(32)}(t), \label{Cij}
\end{eqnarray}
where we introduce
\begin{eqnarray}
\psi^{(11)}(t) &=&  B^2(t) 
\int d \bv{k}
\frac{k_x^2 k_y (k_y - \dot{\gamma} tk_x)}{k^3 k(-t)}
 e^{- \beta(t)k^2}, \nonumber \\
\psi^{(22)}(t) &=&  B^2(t) 
\int d \bv{k}
\frac{k_x^2 k_y (k_y - \dot{\gamma} tk_x)}{ k^4(-t)}
 e^{- \beta(t)k^2}, \nonumber \\
\psi^{(32)}(t) 
&=&  B^2(t) 
\int d \bv{k}
\frac{k_z^2}{ k^2(-t)}
G(\bv{k},t)
 e^{- \beta(t)k^2}
 \label{psi:def}
\end{eqnarray}
with
\begin{eqnarray}
G(\bv{k},t) =
\left[ 
1 + \frac{k_y}{k_\perp}
\left \{ 
  \tan^{-1} \left ( \frac{k_y - \dot{\gamma} t k_x}{k_\perp} \right )
  - \tan^{-1} \left ( \frac{k_y}{k_\perp} \right )
  \right \}
  \right ].
\end{eqnarray}

From (\ref{psi:def}), $\psi^{(11)}(t)$, $\psi^{(22)}(t)$, and $\psi^{(32)}(t)$
obey $t^{-3/2}$ for $t<\dot{\gamma}^{-1}$. 
Hence, we find 
\begin{eqnarray}
C_\eta(t) \propto t^{-3/2}
\end{eqnarray}
for $t<\dot{\gamma}^{-1}$. 

However, $\psi^{(11)}(t)$, $\psi^{(22)}(t)$,
and $\psi^{(32)}(t)$ can be evaluated as
\begin{eqnarray}
\psi^{(11)}(t) &\simeq&  \frac{8 b^{1/2}}{\dot{\gamma} ^3 t^4} c^{(11)}_H, \nonumber \\
\psi^{(22)}(t) &\simeq&  \frac{8 b^{1/2}}{\dot{\gamma} ^3 t^4} c^{(22)}_H, \nonumber \\
\psi^{(32)}(t) &\simeq&  \frac{4 b^{1/2}}{\dot{\gamma} ^3 t^2} c^{(32)}_H, 
\label{psi:SH}
\end{eqnarray}
for $t \gg \dot{\gamma}^{-1}$ in SH.
Here, $c^{(11)}_H$, $c^{(22)}_H$, and $c^{(32)}_H$ are the following constants
\begin{eqnarray}
c^{(11)}_H &=&  
\int d \bv{k}
\frac{k_x^2 k_y (k_y - \sqrt{2} k_x) }
{(k_y^2 + k_z^2)^{3/2} (2 k_x^2 + k_y^2 + k_z^2 - 2\sqrt{2} k_x k_y)^{1/2}}
 e^{- (k^2 - \frac{4 \sqrt{2}}{3} k_x k_y}, \nonumber \\
c^{(22)}_H &=&  
\int d \bv{k}
\frac{k_x^2 k_y (k_y - \sqrt{2} k_x) }
{(2 k_x^2 + k_y^2 + k_z^2 - 2\sqrt{2} k_x k_y)^{2}}
 e^{- (k^2 - \frac{4 \sqrt{2}}{3} k_x k_y}, \nonumber \\
c^{(22)}_H &=&  
\int d \bv{k}
\frac{k_z^2 }
{2 k_x^2 + k_y^2 + k_z^2 - 2\sqrt{2} k_x k_y}
 e^{- (k^2 - \frac{4 \sqrt{2}}{3} k_x k_y} \nonumber \\
& & \times \left[ 
1 + \frac{k_y}{|k_z|}
\left \{ 
  \tan^{-1} \left ( \frac{k_y - \sqrt{2} k_x}{|k_z|} \right )
  - \tan^{-1} \left ( \frac{k_y}{|k_z|} \right )
  \right \}
  \right ].
\end{eqnarray}
Thus, we find that
\begin{eqnarray}
C_\eta(t) \propto \psi^{(32)}(t) \propto t^{-2}
\end{eqnarray}
for $t \gg \dot{\gamma}^{-1}$ in SH.

On the other hand,
$\psi^{(11)}(t)$, $\psi^{(22)}(t)$, 
and $\psi^{(22)}(t)$ behave for $t \gg \dot{\gamma}^{-1}$ as 
\begin{eqnarray}
\psi^{(11)}(t) &\simeq&  \frac{3 \sqrt{3}}{\dot{\gamma}^3  t^{9/2}} c^{(11)}_T, \nonumber \\
\psi^{(22)}(t) &\simeq& \frac{3 \sqrt{3}}{\dot{\gamma}^3   t^{9/2}} c^{(22)}_T, \nonumber \\
\psi^{(32)}(t) &\simeq&  \frac{\sqrt{3}}{\dot{\gamma}   t^{5/2}} c^{(32)}_T.
\label{psi:ST}
\end{eqnarray}
Here $c^{(11)}_T$, $c^{(22)}_T$, and $c^{(32)}_T$ are the constants 
\begin{eqnarray}
c^{(11)}_H &=&  
\int d \bv{k}
\frac{k_x^2 k_y (k_y - \sqrt{3} k_x) }
{(k_y^2 + k_z^2)^{3/2} (3 k_x^2 + k_y^2 + k_z^2 - 2\sqrt{3} k_x k_y)^{1/2}}
 e^{- (k^2 - \sqrt{3} k_x k_y)}, \nonumber \\
c^{(22)}_H &=&  
\int d \bv{k}
\frac{k_x^2 k_y (k_y - \sqrt{3} k_x) }
{(3 k_x^2 + k_y^2 + k_z^2 - 2\sqrt{3} k_x k_y)^{2}}
 e^{- (k^2 - \sqrt{3} k_x k_y)}, \nonumber \\
c^{(22)}_H &=&  
\int d \bv{k}
\frac{k_z^2 }
{3 k_x^2 + k_y^2 + k_z^2 - 2\sqrt{3} k_x k_y}
 e^{- (k^2 -  \sqrt{3} k_x k_y)} \nonumber \\
 & & \times 
\left[ 
1 + \frac{k_y}{|k_z|}
\left \{ 
  \tan^{-1} \left ( \frac{k_y - \sqrt{3} k_x}{|k_z|} \right )
  - \tan^{-1} \left ( \frac{k_y}{|k_z|} \right )
  \right \}
  \right ].
\end{eqnarray}
Thus, we find that
\begin{eqnarray}
C_\eta(t)\propto \psi^{(32)}(t) \propto t^{-5/2}
\end{eqnarray}
for $t \gg \dot{\gamma}^{-1}$ in ST and SG.

Hence, from Eqs. (\ref{Ceq:exp}), (\ref{Cij}), (\ref{psi:SH}) and 
(\ref{psi:ST}), we find that
$C_\eta(t)$ behaves as $t^{-3/2}$ for $t<\dot{\gamma}^{-1}$ 
and $t^{-2}$ for $t \gg \dot{\gamma}^{-1}$ in SH.
$C_\eta(t)$ behaves as $t^{-3/2}$ for $t<\dot{\gamma}^{-1}$ 
and $t^{-5/2}$ for $t \gg \dot{\gamma}^{-1}$ in ST and SG.

\section{Discussion and Conclusion}

In the previous paper \cite{Otsuki07}, we find that
the exponent of the tail in the velocity autocorrelation function $C(t)$
has the crossover from $-3/2$ to $-3$ for SH, 
and $-3/2$ to $-5/2$ for ST and SG.
The exponents for ST and SG are common in $C_\eta(t)$ and $C(t)$.
However, the exponents in $C_\eta(t)$ for SH are 
different from $C(t)$.
This difference is originated from the 
fact that there is $B^2(t)$ in Eq. 
(\ref{psi:def}), but the corresponding part in $C(t)$ is $B(t)$.

In conclusion, we have analytically calculated the behaviors of the
$C_\eta(t)$ in the three dimensional sheared fluids. 
Based on the method developed by 
by Ernst {\it et al.}, we find that $C_\eta(t)$ has the long-time tail
where the exponent has the cross over from $-3/2$ to $-2$ for SH
and $-3/2$ to $-5/2$ for ST and SG.

This work is partially supported by Ministry of Education, Culture, Science and Technology (MEXT), Japan (Grant No. 18540371)
and the Grant-in-Aid for the 21st century COE "Center for Diversity and Universality of Physics" from MEXT, Japan.
One of the authors (M. O.) thanks the Yukawa Foundation for the financial
support.

\end{document}